\documentclass[twocolumn,showpacs,showkeys,a4paper]{revtex4}
\usepackage{amssymb}
\usepackage{graphics}
\usepackage{epsfig}
\usepackage[T1]{fontenc}
\usepackage[latin1]{inputenc}
\begin{document}
\title{ Modified KdV hierarchy : Lax pair representation and bi-Hamiltonian structure } 
\author{Amitava Choudhuri$^\dag$, Benoy Talukdar$^{\ddag}$ and Umapada Das$^*$}
\affiliation{$^\dag$$^\ddag$ Department of Physics, Visva-Bharati University, Santiniketan 731235, India}
\email{binoy123@bsnl.in,  amitava_ch26@yahoo.com}
\affiliation{$^*$\it Abhedananda Mahavidyalaya, Sainthia 731234, India}
\begin{abstract}
We consider equations in the modified KdV (mKdV) hierarchy and make use of the Miura transformation to construct expressions for their Lax pair. We derive a Lagrangian-based approach to study the bi-Hamiltonian structure of the mKdV equations. We also show that the complex modified KdV (cmKdV) equation follows from the action principle to have a Lagrangian representation. This representation not only provides a basis to write the cmKdV equation in the canonical form endowed with an appropriate Poisson structure but also help us construct a semianalytical solution of it. The solution obtained by us may serve as a useful guide for purely numerical routines which are currently being used to solve the cmKdV eqution. 
\end{abstract}
\vskip 0.2 cm
\maketitle
\vskip 0.5 cm
\noindent{\bf 1. Introduction}
\vskip 0.2 cm
Nearly forty years ago Lax {[}1{]} showed that the Korteweg de Vries (KdV) initial value problem for $u=u(x,t)$ given by
$$u_t=u_{3x}-6uu_x\eqno(1)$$
with 
$$u(x,0)={\cal V}(x)\eqno(2)$$
is but one of the infinite family of equations that leave the eigenvalue of the Schr{\"o}dinger equation with the potential  ${\cal V}(x)$ invariant in time. The subscripts of $u$ in $(1)$ denote differentiation with respect to the associated  independent variables. The family of equations discovered by Lax often goes by the name KdV hierarchy and is generated by making use of the recursion operator {[}2{]}
$$\Lambda= \partial_x^2-4u-2u_x\partial_x^{-1}\,\,,\partial_x=\frac{\partial}{\partial x}\eqno(3)$$
in the differential relation
$$u_t=\Lambda^nu_x\,\,,\,\,\,\,n=0,\,1,\,2,\,3,...\,\,\,.\eqno(4)$$
\par The KdV equation $(1)$ is recognized as the solvability condition for the system
$$L\psi=\lambda \psi\eqno(5a)$$
and
$$\partial_t\psi=A\psi\,\,,\partial_t=\frac{\partial}{\partial t}\eqno(5b)$$
with
$$L=-\partial_x^2+u\,\,,\eqno(6a)$$
the so-called Schr{\"o}dinger operator. Here ${\cal A}$ is a third-order linear operator written as 
$$A=4\partial_x^3-3u\partial_x-3\partial_x.u\,\,.\eqno(6b)$$
The existence of the solution $\psi=\psi(\lambda, x, t)$ for every constant $\lambda$ is equivalent to
$$\partial_tL=AL-LA=\left[A,L\right]\,\,.\eqno(7)$$
The result in $(7)$ is called the Lax equation and the operators $L$ and $A$ are called Lax pair {[}1{]}. The Lax pair representation holds good for all equations in the KdV hierarchy. In the context of Lax's method it is often said that $L$ defines the original spectral problem while $A$ represents an auxiliary spectral problem. As one goes along the hierarchy, $L$ remains unchanged but the differential operator associated with the auxiliary spectral problems changes according to  
$$A_n=(4)^n\partial_x^{2n+1}+\sum_{j=1}^n\{a_j\partial_x^{2j-1}+\partial_x^{2j-1}a_j\}\,\,,$$$$n=0,\,1,\,2,\,3,...\,\,\,.\eqno(8)$$
The operator $A_0=\partial_x$ and $A_1$ stands for $A$ in $(6b)$.  It is not always easy to obtain results for other $A_n$'s which generate higher-order KdV equations. The coefficient $a_j$ depends on the solution $u$ and derivatives $u_n(=\frac{\partial^n u}{\partial x^n})$. From $(8)$ it is clear that as $j$ varies, the dimension of $a_j$ changes. Thus $a_j$ should be chosen as a linear combination of power and products of $u$ and $u_n$'s such that the terms in the curly bracket have the right dimension of $\partial_x^{2n+1}$. The constructed expression for $A_n$ will then generate the KdV hierarchy when used in the Lax equation {[}3{]}. On the other hand, one can postulate that for an evolution equation of the form $u_t=K[u]$ the terms in the Fr\'echet derivative  of $K[u]$  contribute additively with unequal weights to form the operator $A_n$ such that $L$ and $A_n$ via (7) reproduces the equations in the hierarchy {[}4{]}. Of course, there should not be any inconsistency in determining the values of the weight factors.
\par   Zakharov and Faddeev {[}5{]} developed the Hamiltonian approach to integrability of nonlinear evolution equations in one spatial and one temporal (1+1) dimensions and, in particular, Gardner {[}6{]} interpreted the KdV equation as a completely integrable Hamiltonian system with $\partial_x$ as the relevant Hamiltonian operator. A significant development in the Hamiltonian theory is due to Magri {[}7{]}, who realized that integrable Hamiltonian systems have an additional structure. They are bi-Hamiltonian i.e. they are Hamiltonian with respect to two different compatible Hamiltonian operators $\partial_x$ and $\left(\partial_x^3-4u\partial_x-2u_x\right)$ such that
$$u_t=\partial_x\left(\frac{\delta {H}_{n}}{\delta u}\right)=\left(\partial_x^3-4u\partial_x-2u_x\right)\left(\frac{\delta {H}_{n-1}}{\delta u}\right)\,\,,$$$$n=1,\,2,\,3 ...\,\,.\eqno(9)$$
Here $H_n=\int {\cal H}_ndx$ with ${\cal H}_n$, the conserved densities for the equations in the KdV hierarchy. These conserved densities generate flows which commute with the KdV flow and as such give rise to an appropriate hierarchy. Traditionally, the expression for ${\cal H}_n$ is constructed using a mathematical formulation that does not make explicit reference to the Lagrangians of the equations in the hierarchy. However, a Lagrangian-based approach can be used to identify ${\cal H}_n$ as the Hamiltonian density of the nth hierarchical equation {[}8{]}.
\par The nonlinear transformation of Miura or the so-called Miura transformation {[}9{]}
$$u=v_x+v^2\,\,,\,\,\,v=v(x,t)\eqno(10)$$
converts the KdV equation into a modified KdV (mKdV) equation
$$v_t=v_{3x}-6v^2v_x\,\,.\eqno(11)$$
This equation differs from the KdV equation only because of its cubic nonlinearity. It has many applicative relevance. For example, mKdV equation has been used to describe acoustic waves in anharmonic lattices and Alfv\'en waves in collisionless plasma. It is of interest to note that the recursion operator $\Lambda$ for the mKdV equation {[}10{]}
$$\Lambda_m= \partial_x^2-4v^2-4v_x\partial_x^{-1}.v\eqno(12)$$
can be identified from $(4)$ and $(10)$. The equation of the mKdV hierarchy can be generated by using the relation
$$v_t=\Lambda_m^nv_x\,\,,\,\,\,\,n=0,\,1,\,2,\,3,...\,\,\,.\eqno(13)$$
It is straightforward to obtain the equations in the mKdV hierarchy from those in the KdV hierarchy by the use of Miura transformation. However, it is a nontrivial problem to derive the Lax representation and construct the bi-Hamiltonian structure of the equations in mKdV hierarchy starting from corresponding results for the KdV equation {[}11{]}. In this work we shall deal with these problems. To derive the Magri structure we shall make use of a Lagrangian-based approach.\par In addition to the above another system of our interest is the complex modified KdV (CMKdV) equation given by
$$v_t=v_{3x}-6|v|^2v_x\,\,,\eqno(14)$$
This equation follows from the third-order nonlinear Schr\"odinger equation via an appropriate variable transformation {[}12{]}. We shall provide a variational formulation of $(14)$ which, on the one hand, allows us to study its canonical structure and, on the other hand, serves as a useful basis to construct an approximate analytical solution in term of a trial function. In this context we note that the numerical routine for solving such equations is quite complicated {[}13{]} and requires the use of Crank Nicolson method for time integration and quintic B-spline function for space integration. We believe the solution presented by us may serve as an initial guide for the more ambitious programmes.\par In $\S 2$ we introduce the equations in the mKdV hierarchy and derive their Lax pair representation. We find that the system of equations follows from the action principle and as such can be obtained from appropriate Lagrangian densities via the so-called Euler-Lagrange equations. The corresponding Hamiltonian densities constitute the  involutive conserved densities of the mKdV equation. We then study the bi-Hamiltonian structure of the mKdV equations. In $\S 3$ we convert $(14)$ to a variational problem and thus obtain a Lagrangian representation for the equation. As a useful application of the Lagrangian density so derived we work out the canonical form {[}5{]} of the cmKdV equation and also construct a solution of it by means of $sech$ trial functions and a Ritz optimization procedure {[}14{]}. Finally, in $\S 4$ we try to summarize our outlook on the present work.  
\vskip 0.3 cm
\noindent{\bf\large 2. mKdV hierarchy}
\vskip 0.3 cm 
\par The equations of the mKdV hierarchy follow from $(13)$ for $n=0,\,1,\,2,\,3...\,.$ We shall construct Lax pair representations of these equations by taking recourse to the use of $(10)$ in $(6a)$ and $(8)$. For these equations we shall use a Lagrangian-based method to obtain the conserved densities which are in involution and generate the so-called mKdV flow. We shall then try to realize the bi-Hamiltonian structure by an appropriate modification of $(9)$ by the use of Miura transform.\\
\\
(a) Lax pair representation\\
\\
From $(6a)$, $(6b)$ and $(10)$ we write
$$L=-\partial_x^2+v^2+v_x\eqno(15a)$$
and
$$A=4\partial_x^3-3(v^2+v_x)\partial_x-3\partial_x.(v^2+v_x)\,\,.\eqno(15b)$$
Using $(15)$ in $(7)$ we get
$$(\partial_x+2v)\left(v_t-v_{3x}+6v^2v_x\right)=0\,\,.\eqno(16)$$
As with $(11)$, $(16)$ gives the mKdV equation. In view of this we shall denote the Lax pair in $(15)$ by $L^m$ and $A^m$ just to indicate that these refer to the mKdV equation. We shall follow this convention for all operators and functions related to the mKdV equation. Consistently with the notation of $(8)$ $A^m (=A)$ stands for $A_1^m$. In close analogy with the case of higher KdV equations the original spectral problem for the mKdV equations characterized by the operator $L^m$ remains unchanged as we go up the hierarchy but the differential operator $A_n^m$'s change with $n$. From $(10)$ and the results given in refs $3$ and $4$ one can calculate the expressions for $A_n^m\,,\,\,\,n=2,\,3,\,4,...\,.$ In the following we present some of our results.
$$A_2^m=16\partial_x^5+(25v_x^2+30v^2v_x+10vv_{2x}+15v^4+5v_{3x})\partial_x-$$$$20(v^2+v_x)\partial_x^3+\partial_x.(25v_x^2+30v^2v_x+10vv_{2x}+15v^4+5v_{3x})-$$$$20\partial_x^3.(v^2+v_x)\,\,,\eqno(17a)$$
$$A_3^m=64\partial_x^7-140(v^4+3v_x^2+2v^2v_x+2vv_{2x}+v_{3x})\partial_x^3+$$$$112(v^2+v_x)\partial_x^5-140\partial_x^3.(v^4+3v_x^2+2v^2v_x+2vv_{2x}+v_{3x})+$$$$112\partial_x^5.(v^2+v_x)+(70v^6+210v^4v_x+1050v^2v_x^2+210v_x^3+$$$$140v^3v_{2x}+840vv_xv_{2x}+721v_{2x}^2+70v^2v_{3x})\partial_x+$$$$(798v_xv_{3x}+182v_xv^4+91v_{5x})\partial_x+$$$$\partial_x.(70v^6+210v^4v_x+1050v^2v_x^2+210v_x^3)$$$$+\partial_x.(140v^3v_{2x}+840vv_xv_{2x}+721v_{2x}^2+70v^2v_{3x}+$$$$798v_xv_{3x}+182v_xv^4+91v_{5x})\eqno(17b)$$
and 
$$A_4^m=256\partial_x^9-255v_{8x}-1794v_{7x}\partial_x-510vv_{7x}-1152v_x\partial_x^7-$$$$1152v^2\partial_x^7-5628v_{6x}\partial_x^2+2100v_xv_{6x}-3588vv_{6x}\partial_x+5670v^2v_{6x}-$$$$4032v_{2x}\partial_x^6-8064vv_x\partial_x^6-10248v_{5x}\partial_x^3-7224v_{2x}v_{5x}-$$$$11256vv_{5x}\partial_x^2-15594v_xv_{5x}\partial_x+18312vv_xv_{5x}+5934v^2v_{5x}\partial_x+$$$$11340v^3v_{5x}-8736v_{3x}\partial_x^5-17472vv_{2x}\partial_x^5-15456v_x^2\partial_x^5+$$$$4032v^2v_x\partial_x^5+2016v^4\partial_x^5-11760v_{4x}\partial_x^4-11760v_{3x}v_{4x}-$$$$20496vv_{4x}\partial_x^3-39204v_{2x}v_{4x}\partial_x+19152vv_{2x}v_{4x}-$$$$43680v_{x}v_{4x}\partial_x^2+12600v^2 v_{4x}\partial_x^2+68670v_x^2 v_{4x}+41100vv_{x}v_{4x}\partial_x+$$$$70224v^2v_{x}v_{4x}+11868v^3v_{4x}\partial_x-210v^4v_{4x}-23520vv_{3x}\partial_x^4-$$$$60240v_xv_{2x}\partial_x^4+10320v^2v_{2x}\partial_x^4+20640vv_x^2\partial_x^4+20640v^3v_x\partial_x^4-$$$$25758v_{3x}^2\partial_x+12180vv_{3x}^2-66864v_xv_{3x}\partial_x^3+15120v^2v_{3x}\partial_x^3-$$$$87360v_{2x}v_{3x}\partial_x^2+170268v_xv_{2x}v_{3x}+69720v_xv_{2x}v_{3x}\partial_x+$$$$130200v^2v_{2x}v_{3x}+75600vv_xv_{3x}\partial_x^2+25200v^3v_{3x}\partial_x^2+$$$$81660v_x^2v_{3x}\partial_x+64596vv_x^2v_{3x}+93336v^2v_xv_{3x}\partial_x-$$$$15960v^3v_xv_{3x}-6300v^4v_{3x}\partial_x-420v^5v_{3x}-50568v_{2x}^2\partial_x^3+$$$$73920vv_xv_{2x}\partial_x^3+28560v_x^3\partial_x^3+68880v^2v_x^2\partial_x^3-5040v^4v_x\partial_x^3-$$$$1680v^6\partial_x^3+19026v_{2x}^2+50400vv_{2x}^2\partial_x^2+114480v_xv_{2x}^2\partial_x+$$$$88452vv_xv_{2x}^2+67272v^2v_{2x}^2\partial_x-15120v^3v_{2x}^2+118440v_x^2v_{2x}\partial_x^2+$$$$161280v^2v_xv_{2x}\partial_x^2-7560v^4v_{2x}\partial_x^2+208488vv_x^2v_{2x}\partial_x+$$$$57960v_x^3v_{2x}-66780v^2v_x^2v_{2x}-60480v^3v_xv_{2x}\partial_x-$$$$27720v^4v_xv_{2x}-12600v^5v_{2x}\partial_x+1260v^6v_{2x}+$$$$85680v v_x^3\partial_x^2-30240v^3v_x^2\partial_x^2-15120v^5v_x\partial_x^2+28518v_x^4\partial_x-$$$$27720vv_x^4-57960v^2v_x^3\partial_x-37800v^3v_x^3-44100v^4v_x^2\partial_x+$$$$7560v^5v_x^2+2520v^6v_x\partial_x+2520v^7v_x+630v^8\partial_x\,\,.\eqno(17c)$$
Using $(15a)$ and $(17a)$ in $(7)$ we get
$$(\partial_x+2v)\{v_t-v_{5x}+40vv_xv_{2x}+10v^2v_{3x}+10v_x^3-30v^4v_x\}=0\,\,.\eqno(18)$$
The expression inside the curly bracket represents the equation obtained from $(13)$ with $n=2$. Results similar to that in $(18)$ hold good for any pair like $[A_n^m\,,\,L^m]$. This observation serves as a useful check on our results for $A_n^m$ with arbitrary values of $n$.\\
\\
(b) Bi-Hamiltonian structure\\
\\
Here we shall demonstrate that the bi-Hamiltonian structure of $(11)$ and all higher-order equations obtained from $(13)$ with $n=2,\,3,\,4...\,.$ We note that a single evolution equation $u_t=P[u]$, $  u\,\,\epsilon\,\,\mathbb{R}$ is never the Euler-Lagrange equation of a variational problem {[}10{]}. One common trick to put a single evolution equation into a variational form is to replace $v$ by a potential function
$$v=-w_x,\,\, w=w(x,t)\,\,.\eqno(19)$$
The function $w$ is often called the Casimir potential. Our expressions for the Lagrangian densities will be written in terms of $w$ and its appropriate derivatives. Hamiltonian densities obtained by use of Legendre map can, however, be expressed in terms of field variable $v(x,t)$ and its derivatives.\par The linear equation obtained from $(13)$ with $n=0$ reads 
$$v_t=v_x\,\,.\eqno(20)$$
From $(19)$ and $(20)$
$$w_{xt}=w_{xx}=P[w_x]\,\,{\rm (say)}\,\,.\eqno(21)$$
Equivalently, 
$$w_t=w_x\,\,.\eqno(22)$$
In writing $(22)$ we have used the boundary condition limit $w(x,t)=0$ as $x\rightarrow \pm\infty$. The self adjointness of $P[w_x]$ ensures the existance of a Lagrangian for $(21)$ and $(22)$. In this case, the Lagrangian density can be constructed using the homotopy formula {[}10{]}
$${\cal L}[\xi]=\int_{0}^{1}\xi P[\lambda \xi]d\lambda\,\,.\eqno(23)$$
From $(23)$ we get
$${\cal L}^m_0=\frac{1}{2}w_tw_x-\frac{1}{2}w_{x}^{2}\,\,,\eqno(24a)$$
The subscript zero is self explanatory. The Hamiltonian density obtained from $(24a)$ is given by
$${\cal H}^m_0=\frac{1}{2}w_x^2=\frac{1}{2}v^2\,\,.\eqno(25a)$$
The Lagrangian and Hamiltonian densities for the mKdV $(n=1)$ and higher-order equations obtained from $(13)$ for $n=2,\,3\,{\rm and}\,4$ are given by
$${\cal L}^m_1=\frac{1}{2}w_tw_x-\frac{1}{2}w_xw_{3x}+\frac{1}{2}w_{x}^{4}\,\,,\eqno(24b)$$
$${\cal H}^m_1=\frac{1}{2}vv_{2x}-\frac{1}{2}v^4\,\,,\eqno(25b)$$
$${\cal L}^m_2=\frac{1}{2}w_tw_x-\frac{1}{2}w_{3x}^2-w_x^6-5w_x^2w_{2x}^2\,\,,\eqno(24c)$$
$${\cal H}^m_2=\frac{1}{2}v_{2x}^2+v^6+5v^2v_{x}^2\,\,,\eqno(25c)$$
$${\cal L}^m_3=\frac{1}{2}w_tw_x-\frac{1}{2}w_xw_{7x}+\frac{7}{2}w_x^3w_{5x}+14w_x^2w_{2x}w_{4x}+$$$$\frac{21}{2}w_x^2w_{3x}^2+\frac{35}{2}w_xw_{2x}^2w_{3x}-\frac{35}{3}w_x^5w_{3x}-\frac{70}{3}w_x^4w_{2x}^2+\frac{5}{2}w_x^8\,\,,\eqno(24d)$$
$${\cal H}^m_3=\frac{1}{2}vv_{6x}-\frac{7}{2}v^3v_{4x}-14v^2v_xv_{3x}-\frac{21}{2}v^2v_{2x}^2-$$$$\frac{35}{2}vv_x^2v_{2x}+\frac{35}{3}v^5v_{2x}-\frac{70}{3}v^4v_x^2-\frac{5}{2}v^8\eqno(25d)$$
and
$${\cal L}^m_4=\frac{1}{2}w_tw_x-\frac{1}{2}w_xw_{9x}+\frac{9}{2}w_x^3w_{7x}+27w_x^2w_{2x}w_{6x}+$$$$57w_x^2w_{3x}w_{5x}+\frac{105}{2}w_xw_{2x}^2w_{5x}-21w_x^5w_{5x}+\frac{69}{2}w_x^2w_{4x}^2+$$$$189w_xw_{2x}w_{3x}w_{4x}-168w_x^4w_{2x}w_{4x}+\frac{91}{2}w_xw_{3x}^3-126w_x^4w_{3x}^2$$$$-518w_x^3w_{2x}^2w_{3x}+\frac{105}{2}w_x^7w_{3x}-133w_x^2w_{2x}^4+\frac{315}{2}w_x^6w_{2x}^2-7w_x^{10}\eqno(24e)$$
$${\cal H}_4^m=\frac{1}{2}vv_{8x}-\frac{9}{2}v^3v_{6x}-27v^2v_xv_{5x}-57v^2v_{2x}v_{4x}-\frac{105}{2}vv_{x}^2v_{4x}$$$$+21v^5v_{4x}-\frac{69}{2}v^2v_{3x}^2-189vv_{x}v_{2x}v_{3x}+168v^4v_{x}v_{3x}-\frac{91}{2}vv_{2x}^3$$$$+126v^4v_{2x}^2+518v^3v_{x}^2v_{2x}-\frac{105}{2}v^7v_{2x}+$$$$133v^2v_{x}^4-\frac{315}{2}v^6v_{x}^2+7v^{10}\,\,.\eqno(25e)$$
Results of ${\cal H}_n^m$'s for still higher values of $n$ can be obtained in a similar manner. As a useful check on our expressions, one can verify that these results are in exact agreement with those obtained by the application of Miura transformation on the well known conserved densities of the KdV equations.\par The bi-Hamiltonian structure of equations in the mKdV hierarchy can easily be verified by using our Hamiltonian functionals in 
$$v_t=\partial_x\left(\frac{\delta {H}_{n}^m}{\delta v}\right)={\cal E}\left(\frac{\delta {H}^m_{n-1}}{\delta v}\right)\,\,,\,\,\,\,\,\,n=1,\,2,\,3 ...\,\,,\eqno(26)$$
where
$${\cal E}=\left(\partial_x^3-4v^2\partial_x-4v_x\partial_x^{-1}.v\partial_x\right)\,\,.\eqno(27)$$
The first Hamiltonian operator $\partial_x$ in $(26)$ is the same as that in $(9)$ while the second has been obtained from {[}10{]}
$${\cal E}={\Lambda_m}{\partial_x}\,\,.\eqno(28)$$
The operators $\partial_x$ and ${\cal E}$ are skew symmetric and satisfy the Jacobi-identity. Thus they constitute two compatible Hamiltonian operators such that all equations obtained from $(13)$ are integrable in the Liouville's sense {[}7{]}.
\vskip 0.3 cm
\noindent{\bf\large 3. cmKdV equation}
\vskip 0.3 cm
The complex mKdV equation in $(14)$ can be restated as a variational problem given by
$$\delta\int \int{\cal L}\left(v,\,v^*,\,v_x,\,v_x^*,\,v_{3x},\,v_{3x}^*,\,v_t,\,v_t^*,\,x,\,t\right)dx\,dt=0\eqno(29)$$
with the Lagrangian density written as
$${\cal L}={1\over 2}\left(v^*v_t-vv_t^*\right)-{1\over 2}\left(v^*v_{3x}-vv_{3x}^*\right)+{3\over 2}vv^*\left(v^*v_{x}-vv_{x}^*\right)\,\,.\eqno(30)$$
The Euler-Lagrange equations corresponding to $(29)$ are 
$$\frac{d}{dt}\left(\frac{\partial{\cal L}}{\partial v_t}\right)-\frac{\delta {\cal L}}{\delta v}=0  \eqno(31a)$$
and
$$\frac{d}{dt}\left(\frac{\partial{\cal L}}{\partial v_t^*}\right)-\frac{\delta {\cal L}}{\delta v^*}=0  \eqno(31b)$$
with the variational derivative
$$\frac{\delta}{\delta \psi}=\sum_{n\geq 0}^3(-\partial_x)^n\frac{\partial}{\partial \psi_n}\,\,.\eqno(32)$$ 
Here
$$\psi_n=(\partial_x)^n\psi\,\,.\eqno(33)$$
It is easy to verify that $(30)$ and $(31b)$ give the cmKdV equation while $(30)$ and $(31a)$ give the corresponding complex conjugate equation. The Hamiltonian corresponding to the Lagrangian density $(30)$ is given by 
$$H=\int{\cal H} dx\eqno(34)$$
with the Hamiltonian density
$${\cal H} ={1\over 2}\left(v^*v_{3x}-vv_{3x}^*\right)-{3\over 2}vv^*\left(v^*v_{x}-vv_{x}^*\right)\,\,.\eqno(35)$$
In order to show that $(14)$ is a Hamiltonian system we will have to write it and its complex conjugate in two different forms
$$v_t=\{v^*(x)\,,\,H(y)\}\eqno(36)$$
and 
$$v^*_t=-\{v(x)\,,\,H(y)\}\,\,.\eqno(37)$$
We have already found an expression for the Hamiltonian. Thus our task is to look for fundamental Poisson bracket relation for the field variables that reduce $(36)$ to the cmKdV equation and $(37)$ to the complex conjugate one. One can easily check that the required Poisson bracket relations are given by 
$$\{v(x)\,,\,v(y)\}=\{v^*(x)\,,\,v^*(y)\}=\delta(x-y)\,\,.\eqno(38)$$
The relations $(36)$ and $(37)$ can be written in the symplectic form
$${\bf \eta}_t={\bf J}\frac{\delta {\cal H}}{\delta {\bf \eta}} \,\,\,,\,\,\,\,
\eta =\left(\begin{array}{c}
v\\
v^*
\end{array}\right)\eqno(39)$$
with ${\bf J}=\left(\begin{array}{cc}
0&1\\
-1&0
\end{array}\right)\,\,,$ a skew-adjoint matrix as the Hamiltonian operator.\par Equation $(14)$ arises in a number of applicative contexts including the nonlinear evolution of plasma waves {[}15{]}. To our knowledge there is no well-defined spectral problem that can easily be used to solve the cmKdV equation in terms of known transcendental functions. But a number of works has been envisaged to obtain the solitory-waves and/or soliton solutions of this equation. See, for example, {[}13{]} and references therein. Here we are interested to provide an accurate approximation solution of $(14)$ by supplementing the Lagrangian density in $(30)$ with $sech$ trial functions and a Ritz optimization procedure. We have chosen to work with the trial function written as
$$v(x,\,t)=a(t)sech\left[(x-y(t))/w(t)\right]\times$$$$e^{\left[i\left(q(t)+r(t)(x-y(t))+\frac{b(t)}{2w(t)}(x-y(t))^2\right)\right]}\,\,.\eqno(40)$$
Here the parameters $a$, $y$ and $w$ are related to the three lowest-order moments of the $v$ envelope and  represent, respectively, its amplitude, central position and width. The other parameters $q$, $r$ and $b$ stand for the phase, velocity (center of mass) and frequency chirp. Understandably, these parameters will all vary with time $t$. Using $(40)$ in $(30)$ we get 
$${\cal L}_s=\sum_{i=1}^3{\cal L}_s^{(i)}\,\,,\eqno(41)$$
where
$${\cal L}_s^{(1)}={1\over 2}\left(\frac{x-y}{w}\right)^2 a^2bw \,sech^2\left(\frac{x-y}{w}\right)+$$$$a^2 r \dot y \,sech^2\left(\frac{x-y}{w}\right)-a^2 \dot q \,sech^2\left(\frac{x-y}{w}\right)-$$$${1\over 2}\left(\frac{x-y}{w}\right)^2 a^2\dot bw \,sech^2\left(\frac{x-y}{w}\right)\,\,,\eqno(42a)$$

$${\cal L}_s^{(2)}=\frac{3a^2 r}{w^2}\,sech^2\left(\frac{x-y}{w}\right)\,tanh^2\left(\frac{x-y}{w}\right)-$$$$\frac{3a^2 r}{w^2}\,sech^4\left(\frac{x-y}{w}\right)-a^2r^3\,sech^2\left(\frac{x-y}{w}\right)-$$$$3\left(\frac{x-y}{w}\right)^2a^2b^2r\,sech^2\left(\frac{x-y}{w}\right)\eqno(42b)$$
and 
$${\cal L}_s^{(3)}=-3 a^4 r \,sech^4\left(\frac{x-y}{w}\right)\,\,.\eqno(42c)$$
Here the dots stand for derivative with respect to $t$. The subscript $s$ on ${\cal L}$ merely indicates that we have inserted the $sech$ ansatz for $v(x,\,t)$ into the Lagrangian density. In terms of $(41)$ the variational principle $(29)$ leads to
$$\delta \int<L>\,dt=0\,\,,\eqno(43)$$
with the averaged effective Lagrangian
$$<L>=\int_\infty^\infty{\cal L}_sdx\,\,.\eqno(44)$$
The result for $<L>$ is given by
$$<L>=-2wa^2r^3-4wa^4r-{\pi^2\over2} a^2b^2 r w-\frac{2a^2 r}{w}-$$$${\pi^2\over12}w^2a^2\dot b-2 a^2w\dot q+2 a^2 w r\dot y+{\pi^2\over12}a^2 b w \dot w\,\,.\eqno(45)$$
The reduced variational principle expressed by $(43)$ results in a set of coupled ordinary differential equations for the parameters of our trial function. From the vanishing condition of the variationals
$$\frac{\delta <L>}{\delta q}\,\,,\,\,\,\frac{\delta <L>}{\delta a}\,\,,\,\,\,\frac{\delta <L>}{\delta y}\,\,,\,\,\,\frac{\delta <L>}{\delta w}\,\,,\,\,\,\frac{\delta <L>}{\delta r}$$$$\,\,\,\,{\rm and}\,\,\,\,\frac{\delta <L>}{\delta b}$$
we obtain
$$\frac{d}{dt}\left(2a^2w\right)=0\,\,,\eqno(46a)$$
$$-4aw r^3-16w a^3r-\pi^2 a b^2 r w-\frac{4ar}{w}-{\pi^2\over6}w^2 a\dot b-4aw\dot q+$$$$4a w r\dot y+{\pi^2\over6}a b w \dot w=0\,\,,\eqno(46b)$$
$$-\frac{d}{dt}\left(2a^2wr\right)=0\,\,,\eqno(46c)$$
$$-2a^2r^3-4a^4r-{\pi^2\over2} a^2b^2 r+\frac{2a^2 r}{w^2}-{\pi^2\over6}wa^2\dot b-2 a^2\dot q+$$$$2 a^2  r\dot y+{\pi^2\over12}a^2 b  \dot w-\frac{d}{dt}\left({\pi^2\over12}a^2 b w\right)=0\,\,,\eqno(46d)$$
$$-6wa^2r^2-4wa^4-{\pi^2\over2} a^2b^2 w-\frac{2a^2}{w}+2 a^2 w \dot y=0\eqno(46e)$$
and
$$-\pi^2a^2brw+{\pi^2\over12}a^2w+\frac{d}{dt}\left({\pi^2\over12}w^2a^2\right)=0\,\,.\eqno(46f)$$
Equations in $(46)$ can be used to write
$$\!\!\!\!\!\!\!\!\!\!\!\!\!\!\!\!\!\!\!\!\!\!\!a^2w={\rm constant}=E_0\,\,,\eqno(47a)$$
$$\!\!\!\!\!\!\!\!\!\!\!\!\!\!\!\!\!\!\!\!\!\!\!\!\!\!\!\!\!\!\!\!\!\!\!\!\!\!\!\!\!\!\!r={\rm constant}\,\,,\eqno(47b)$$
$$\!\!\!\!\!\!\!\!\!\!\!\!\!\!\!\!\!\!\!\!\!\!\!\!\!\!\!\!\!\!\!\!\!\!\!\!\!\!\!\!\!\frac{da}{dt}=-\frac{3abr}{w}\,\,,\eqno(47c)$$
$$\frac{dy}{dt}=3r^2+2a^2+\frac{\pi^2}{4}b^2+\frac{1}{w^2}\,\,,\eqno(47d)$$
$$\!\!\!\!\!\!\!\!\!\!\!\!\!\!\!\!\!\!\!\!\!\!\!\!\!\!\!\!\!\!\!\!\!\!\!\!\!\!\!\!\!\!\!\!\!\!\!\!\!\!\!\!\!\frac{dw}{dt}=6br\eqno(47e)$$
and
$$\!\!\!\!\!\!\!\!\!\!\!\!\!\!\!\!\!\!\!\!\frac{db}{dt}=\frac{24 r}{\pi^2}\frac{a^2}{w}+\frac{24 r}{\pi^2}\frac{1}{w^3}\,\,.\eqno(47f)$$
Equation $(47a)$ expresses the variational version of the energy conservation law {[}16{]} while $(47b)$ states that the center of mass of the solution of $(14)$ moves with a constant velocity. For a given values of $r$, the set of coupled ordinary differential equations $(47c)$-$(47f)$ can easily be solved numerically. Note that knowledge of $a(t)$, $y(t)$ and $w(t)$ can be used to study the $|v(x,\,t)|$ as functions of $x$ and $t$. We worked with the initial conditions $a(0)=1$, $b(0)=0$, $y(0)=0$, $w(0)=1$ and solved these equations using the fourth-order Runge-Kutta method {[}17{]}.
\begin{figure}
\includegraphics[width=0.80\columnwidth,angle=0]{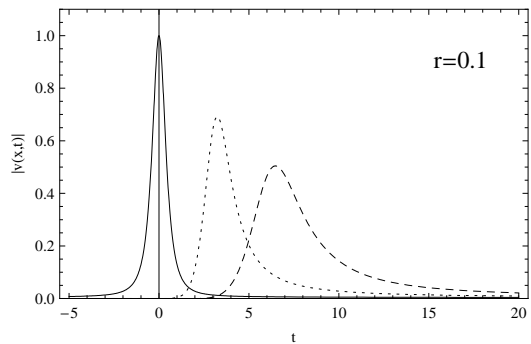}
\hskip 0.5 cm\caption{$|v(x,t)|$ as a function of $t$ for three different \\ values of $x$. Here $r=0.1$.}
\end{figure}
\begin{figure}
\includegraphics[width=0.80\columnwidth,angle=0]{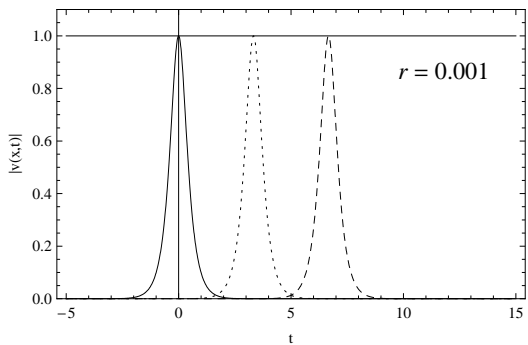}
\caption{Same as that in FIG. $1$ but with $r=0.001$}
\end{figure}
First we take $r=0.1$ and plot in FIG.$1$, $|v(x,\,t)|$ as a function of $t$ for three different values of $x$, namely, $x= 0$ (solid curve), $x=10 $ (dotted curve) and $x= 20$ (dashed curve). From these curves it is clear that as $x$ increases $|v(x,\,t)|$ decreases rapidly. This implies that for our chosen value of the velocity we have decaying solitory wave solution. We have verified that for still higher values of $r$ the solutions decay more rapidly. In FIG. $2$ we present a similar plot of $|v(x,\,t)|$ for $r=0.001$. Interestingly, $|v(x,\,t)|$  remains unchanged as $x$ increases. Thus one can infer that the solutions of $(14)$ for small values of $r$ behaves like solitons. 
\vskip 0.5 cm
\noindent{\bf\large 4. Conclusion}
\vskip 0.3 cm
The nonlinear transformation of Miura or the so-called Miura transformation is an aid to obtain the modified KdV (mKdV) equation from the KdV equation. We find that this transformation also provides an effective way to construct expressions for Lax pairs of all equations in the mKdV hierarchy. As with the KdV equations the bi-hamiltonian structure of the mKdV equations are traditionally studied using involutive set of conserved Hamiltonian densities without explicit reference to their Lagrangians. We derive a Lagrangian-based approach to realize the bi-Hamiltonian structure. \par In close analogy with the mKdV equation, the cmKdV equation in $(14)$ also follows from Hamilton's variational principle provided the action functional is made to vanish for simultaneous variations of both $v$ and $v^*$. In this case the Lagrangian density is a function of $v$, $v^*$ and their appropriate time and space derivatives. We could use the Hamiltonian corresponding to this Lagrangian density to write the cmKdV equation in the canonical form {[}5{]} with an appropriate Poisson structure. As an added realizm we demonstrate that the Lagrangian density constitutes a basis to derive a semianalytical solution of $(14)$. We achieve this by taking recourse to the use of $sech$ trial functions to define a reduced variational problem which in conjunction with the Ritz optimization procedure could yield an uncomplicated solution of the cmKdV equation. There exist some sophisticated numerical routines {[}13,15{]} to solve the equation. However, we feel that the variational approach sought by us will serve as a complementary tool towards understanding the properties of solitory wave- and/or soliton-solutions of the complex modified KdV equation.\\
\\
{\bf Acknowledgements} \\
This work is supported by the University Grants Commission, Government of India, through grant No. F.32-39/2006(SR).
\vskip 0.5 cm
\noindent {\bf References}
\vskip 0.5 cm
{[}1{]}  P. D. Lax, Commun. Pure  Appl. Math., {\bf 21}, 467 (1968).\\

{[}2{]}  F. Calogero and A. Degasperis, Spectral Transform and Soliton (New York: North-Holland Publising Company, 1982).  \\
 
{[}3{]}  S. Chakraborti, J. Pal, J. Shamanna and B. Talukdar, Czech. J.  Phys. {\bf 52}, 853 (2003).\\

{[}4{]} A. Choudhuri, B. Talukdar and S. B. Datta, Z. Naturforsch {\bf 61a}, 7 (2006). \\

{[}5{]} V. E. Zakharov and L. D. Faddeev, Funct. Anal. Appl. {\bf 5}, 18 (1971).\\

{[}6{]} C. S. Gardner, J. Math. Phys. {\bf 12}, 1548 (1971).\\

{[}7{]} F. Magri, J. Math. Phys. {\bf 19}, 1156 (1978).\\

{[}8{]} S. Ghosh, B. Talukdar and J. Shamanna, Czech. J. Phys.  {\bf 53}, 425 (2003).\\

{[}9{]} R. M. Miura, J. Math. Phys. {\bf 9}, 1202 (1968).\\

{[}10{]} P. J. Olver, Application of Lie Groups to Differential Equation, (New York: Springer-Verlag, 1993).\\

{[}11{]} K. Toda, Proc. Inst. Math. of NAS, Ukraine {\bf 43}, 377 (2002).\\

{[}12{]} J. Yang, Phys. Rev. Lett. {\bf 91} 143903 (2003).\\

{[}13{]} M. S. Ismail  {\it Commun. Nonlinear Science and Numerical Simulation} doi: 10.1016, (2008); D. Irk and I Dag, Phys. Scr. {\bf 77}, 065001 (2008).\\

{[}14{]} D. Anderson, Phys. Rev. A. {\bf 27}, 3135 (1983).\\

{[}15{]} G. M. Muslu and H. A. Erbay, Computers and Mathematics with Applications {\bf 45}, 503 (2003).\\

{[}16{]} B. A. Malomed, Progr. Optics {\bf 43}, 69 (2002).\\

{[}17{]} J. B. Scarborough, Numerical Mathematical Analysis, Oxford and IBH Publishing C., New Delhi, 1971.
\end{document}